\documentclass{iaus}
\usepackage{graphicx}

\title[Be in IC 4651] %% give here short title %%
{Beryllium abundances along the evolutionary sequence of the open cluster IC 4651}

\author[Smiljanic et al.]   %% give here short author list %%
{Rodolfo Smiljanic$^{1,2}$, 
L. Pasquini$^2$, 
C. Charbonnel$^{3,4}$,
 \and N. Lagarde$^3$}

\affiliation{$^1$IAG, University of S\~ao Paulo, Brazil, 
$^2$ ESO, Germany, \\ email: {\tt rsmiljan@eso.org} \\[\affilskip] 
$^3$ Geneva Observatory, Switzerland, 
$^4$ LATT, CNRS, Universit\'e de Toulouse, France
}

\pubyear{2010}
\volume{268}  %% insert here IAU Symposium No.
\jname{Light elements in the Universe}
\editors{C. Charbonnel, M. Tosi, F. Primas \& C. Chiappini, eds.}
\begin{document}

\maketitle

\begin{abstract}
The simultaneous investigation of Li and Be in stars is a powerful tool in 
the study of the evolutionary mixing processes. Here, we present beryllium abundances in 
stars along the whole evolutionary sequence of the open cluster IC 4651. This cluster has a 
metallicity of [Fe/H] = +0.11 and an age of 1.2 or 1.7 Gyr. Abundances have been determined from high-resolution, 
high signal-to-noise UVES spectra using spectrum synthesis and model atmospheres. Lithium 
abundances for the same stars were determined in a previous work. Confirming previous results, we find that 
the Li dip is also a Be dip. For post-main-sequence stars, the Be dilution starts earlier within 
the Hertzsprung gap than expected from classical predictions, as does the Li dilution. Theoretical hydrodynamical 
models are able to reproduce well all the observed features.

\end{abstract}

\firstsection % if your document starts with a section,
\section{Introduction}

In contradiction with standard stellar evolution models, where convection is the only mechanism 
driving mixing episodes, field and cluster F- and early G-type stars (including the Sun) deplete Li 
abundances during the main sequence \cite[(Lambert \& Reddy 2004; Sestito \& Randich 2005, and references therein)]{LR04,SR05} 
Different physical mechanisms have been proposed to explain these observations: atomic diffusion, mass loss, rotation-induced 
mixing, internal gravity waves, or combinations of these (see \cite[Charbonnel \& Talon 2008]{CT08} and references therein). 

As Li and Be burn at different temperatures (2.5 x 10$^{6}$ K for Li and 3.5 x 10$^{6}$ K for Be), i.e. at different depths in the 
stellar interior, they help in constraining the transport mechanisms by performing a stellar tomography. In the study of mixing 
processes as a function of mass and evolutionary status cluster stars are ideal because they have well defined masses and 
share the same age and initial chemical composition. We derived Be abundances along the whole evolutionary sequence of the open cluster IC 4651, including 
solar-type, Li-dip, turn-off, subgiant, and red giant stars. With these new results, we investigate in detail the mixing processes in different 
stellar masses.

\section{Discussion}

The sample analyzed here is composed of 22 stars; 21 with atmospheric parameters and Li abundances from 
\cite[Pasquini et al. (2004)]{Pas04} and 1 from \cite[Randich et al. (2002)]{Ran02}. The spectra 
have 40 $\leq$ S/N $\leq$ 100 (per resolution element) and R $\sim$ 45000. 

Be abundances were determined using synthetic spectra and the same codes and line lists used 
in \cite[Smiljanic et al. (2009a)]{Sm09a}. As some of the sample stars are fast rotators, we first carefully 
modeled the slow-rotating stars and tested the effects in the abundances of artificially 
broadening the spectra (see \cite[Smiljanic et al. 2009b]{Sm09b} for details). Beryllium was detected in all 
the sample stars except for the giants.

\begin{figure}[t]
% \vspace*{-2.0 cm}
\begin{center}
 \includegraphics[width=4.6in]{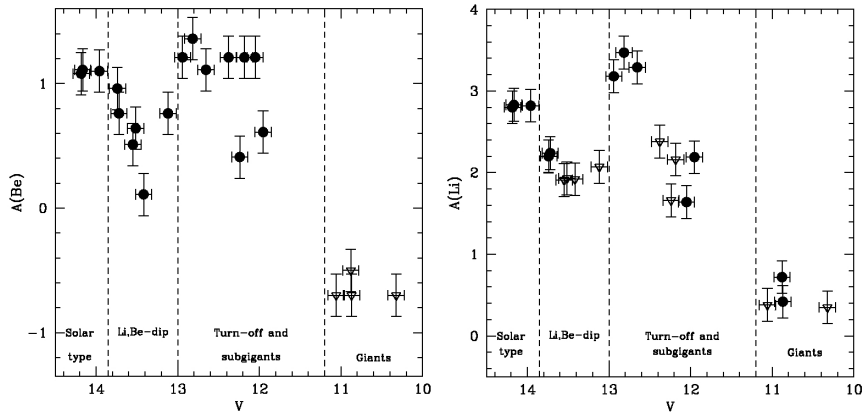} 
% \vspace*{-1.0 cm}
 \caption{Abundances of Be (left panel) and Li (right panel) as a function of the V magnitude. Detections 
 are shown as full circles and upper limits as open triangles.}
   \label{fig1}
\end{center}
\end{figure}

New evolutionary models for stars on the hot side of the dip, including atomic diffusion, meridional circulation, 
and shear turbulence, were calculated with STAREVOL V3.1 by Lagarde \& Charbonnel (in preparation, see also Charbonnel \& Lagarde this volume) for a 
range of stellar masses and initial rotation velocities.  For stars on the cool side of the Li dip we use the 
1.2 M$_{\odot}$ model computed by \cite[Talon \& Charbonnel (2005)]{TC05} which has an initial rotation velocity of 50 km s$^{-1}$.

%, that have deeper 
%convective envelopes and where internal gravity waves participate on the extraction of angular momentum 
%with meridional circulation and shear turbulence. In this case,

Beryllium abundances are found to follow closely the behavior of the Li abundances (Fig. 1). In a 
sequence of increasing mass we have first the coolest main-sequence stars that do not present a Be 
abundance dispersion. This is expected to be due to the impact of internal gravity waves. After that, a well-defined 
Be dip is seen. This confirms previous results that the Li dip is also a Be dip 
\cite[(Boesgaard \& King 2002, Boesgaard et al. 2004)]{BK02,Boe04}. For post-main-sequence stars 
we confirm that Be dilution starts earlier than the expected classically. The Be abundances also  present a significant dispersion. 

The dispersion of Li and Be abundances on the blue side of the dip and in evolved stars is very well 
explained by the models when accounting for a dispersion in the initial values of the stellar rotational velocities. 
The models reproduce all the Li and Be features along the CMD of IC 4651. The success in 
explaining the Li and Be abundances along the whole evolutionary sequence shows that important 
steps have been taken towards the proper understanding of the physical mechanisms acting during 
the stellar evolution.

\end{document}